\documentclass[epj,D]{svjour}
\usepackage{graphics}
\begin{document}
\title{Universally-composable finite-key analysis for efficient four-intensity decoy-state quantum key distribution}
\author{Haodong Jiang \inst{1}, Ming Gao \inst{1, }\thanks{E-mail: gaoming.zhengzhou@gmail.com}, Bao Yan \inst{1}, Weilong Wang \inst{1} \and Zhi Ma \inst{1}
%
}                     
%
%
\institute{State Key Laboratory of Mathematical Engineering and Advanced Computing, Zhengzhou, Henan, China}
\date{Received: XXX/ Revised version: XXX}
%

\abstract{
We propose an efficient four-intensity decoy-state BB84 protocol and derive concise security bounds for this protocol with the universally composable finite-key analysis method.
Comparing with the efficient three-intensity protocol, we find that our efficient four-intensity protocol can increase the secret key rate by at least $30\%$. Particularly, this increasing rate of secret key rate will be raised as the transmission distance increases.
At a large transmission distance, our efficient four-intensity protocol can improve the performance of quantum key distribution profoundly.
} 

\maketitle
\section{Introduction}
Combining with one time pad \cite{shannon1949communication}, quantum key distribution (QKD) \cite{bennett1984quantum,ekert1991quantum} can offer a private communication with an information-theoretical security \cite{lo1999unconditional,shor2000simple,mayers2001unconditional,renner2005information}.
In practical QKD implementations, a weak pulsed laser source is utilized in place of an ideal single-photon source. To deal with the security vulnerability coming from the multi-photon components of emitted laser pulses \cite{brassard2000limitations,pns2002quantum}, decoy-state method is proposed \cite{hwang2003quantum,wang2005beating,lo2005decoy}. With this method, the secure single-photon contribution can be estimated effectively.

In asymptotic setting (with infinitely long keys), the security of decoy-state QKD is analyzed \cite{wang2005beating,lo2005decoy}.
In the case of finite-length keys, security bounds against general attacks are first derived in Ref. \cite{hayashi2014security}.
Subsequently, Ref. \cite{lim2014concise} derives concise and tight finite-key security bounds for efficient three-intensity decoy-state protocol by combining the recent security proof technique \cite{tomamichel2011uncertainty,tomamichel2012tight} with a finite-key analysis for decoy-state method.
The simulation results show that these bounds are relatively tight.

Four-intensity decoy-state protocols are researched in \cite{wang2005decoy,hayashi2007general,zhou2014tightened,yu2015decoy}.
However, in their works, the four different intensities are mainly used to obtain a tighter estimation formula for single-photon error rate \cite{hayashi2007general,zhou2014tightened} and
the secret key rates are calculated
in asymptotic setting.
For practical QKD implementations, the effects due to finite-length keys should be considered, e.g., statistical fluctuation \cite{hayashi2014security,lim2014concise,cai2009finite}.
Thus, in finite-key setting, statistical fluctuations of four different measurement values (the numbers of quantum bit errors) should be taken into account when four intensities are all utilized for one estimation formula of single-photon error rate \cite{hayashi2007general,zhou2014tightened}, which may in contrary bring a lower secret key rate especially when the transmission distance is large.

Here, we propose an efficient four-intensity decoy-state QKD protocol with biased basis choice.
Unlike previous four-intensity protocols, in this protocol, the basis choice is biased and the estimation method for single-photon contribution is the same with the widely used one \cite{lim2014concise,ma2005practical}.
Additionally, different from efficient three-intensity protocols \cite{lim2014concise,wei2013decoy,lucamarini2013efficient}, the intensities and the bases in our protocol are independent except the lowest intensity.
More specifically, in our protocol, $Z$ basis is used for key generation where three different intensities are utilized, and $X$ basis is used for testing where two different intensities are used \cite{Lo2005efficient}.
The two higher intensities in \(Z\) basis (except lowest intensity) are independent from the higher intensity in $X$ basis.
Compared with efficient three-intensity protocols, the intensities in our protocol can be freely optimized to increase the detected pulses in two bases (for a fixed number of sent pulses $N$) and decrease the statistical deviations caused by finite-length key.

 Using the universally composable finite-key analysis method \cite{lim2014concise}, we derive concise security bounds for our efficient four-intensity protocol.
With these bounds and system parameters in Ref. \cite{lim2014concise}, we perform some numerical simulations with full parameter optimization.
When the number of sent pulses $N$ is \(10^9\), compared with efficient three-intensity protocol \cite{lim2014concise}, our protocol can increase the secret key rate by at least \(30\%\).
Particularly, this increasing rate of secret key rate will be raised with the increasing transmission distance.
\section{Protocol description}
In this paper, we consider the efficient BB84 protocol \cite{Lo2005efficient}, i.e., the basis choice is biased.
Our protocol is based on the transmission of phase-randomized laser pulses, and uses four different-intensity setting.
Then, we describe our efficient four-intensity protocol in detail.

\textit{1. Preparation and measurement.}
Alice sends four different kinds of weak laser pulses with intensities \(\omega ,{\rm{ }}{\upsilon _1},{\rm{ }}{\upsilon _2},\mu {\kern 1pt} {\kern 1pt} {\kern 1pt} {\kern 1pt} {\kern 1pt}\) \((\mu  > {\upsilon _1} + \omega ,{\upsilon _1} > \omega  \ge 0,{\upsilon _2} > \omega  \ge 0)\),
with probabilities, \({P_\omega }\), \({P_{{\upsilon _1}}}\), \({P_{{\upsilon _2}}}\) and \({P_\mu }\) (\(P_\omega+P_{{\upsilon _1}}+P_{{\upsilon _2}}+P_\mu=1 \)), respectively.
Specially, the pulses with intensities (\({\upsilon _1}\) and $\mu$) are all prepared in $Z$ basis, the pulses with intensity \({\upsilon _2}\) are all prepared in $X$ basis, and the pulses with the lowest intensity \(\omega\) are prepared in $Z(X)$ basis with probability \({P_{Z|\omega }}({P_{X|\omega }})\) (\(P_{Z|\omega }+P_{X|\omega }=1\)).
Bob chooses $Z(X)$ basis to perform the measurement with probability \(P_Z{(P_X)}\) ($P_Z+P_X=1$).

\textit{2. Basis reconciliation and parameter estimation.}
Alice and Bob announce their basis choices over an authenticated public channel, and accomplish the sifting by reserving the detected signals with the same basis and discarding the others.
After this procedure, Alice and Bob share two bit strings with lengths \({n_Z}\) and ${n_X}$ corresponding to two bases.
Then, Alice announces the intensity information.
Based on that, the shared bit string in \(Z\) basis can be divided into three substrings with lengths,
\({n_{Z,\nu }}\) (\(\nu \in \{ \mu,{\upsilon _1},\omega \}\)), where \({n_Z} = \sum\limits_{\nu  \in \{ \mu ,{\upsilon _1},\omega \} } {{n_{Z,\nu }}} \).
As for \(X\) basis, Alice and Bob need to announce their bit strings.
They compare them and obtain the numbers of bit error, \({m_{X,k}}\) (\(k \in \{ {\upsilon _2},\omega \}\)).
With these \({n_{Z,\nu }}\) and \({m_{X,k}}\), they can calculate the number of vacuum events \(s_{Z,0}\) [Eq. (\ref{Eq:sZ0})], the number of single-photon events \(s_{Z,1}\) [Eq. (\ref{Eq:sZ1})] and the \textit{phase error rate} \(e_{1}^{PZ}\) [Eq. (\ref{Eq:e1PZ})] associated with single-photon events in \(Z\) basis.

\textit{3. Error correction and privacy amplification.}
In the error-correction step, we assume that the error rate \(E_Z\) is predetermined and at most \(\lambda _{EC}=f{n_z}H({E_z})\) is revealed, where \(f\) is error-correction efficiency; \(H(x)=- x{\log _2}(x) - (1 - x){\log _2}(1 - x)\) is the binary Shannon entropy function.
Next, an error verification is performed to ensure that Alice and Bob share a pair of identical keys.
\({\varepsilon _{cor}}\) is the probability that a pair of nonidentical keys pass this error-verification step.
Finally, they perform the privacy amplification to extract the \(\varepsilon _{sec}\)-secret keys with length \(l\) [Eq.~\ref{Eq:l}].

\section{Security bounds for efficient four-intensity protocol}
Following the finite-key security analysis in Ref. \cite{lim2014concise}, the length of \(\varepsilon_{sec}\)-secret key \(l\) is given
by
\begin{eqnarray}\label{Eq:l}
l =&
\lfloor{s_{Z,0}} + {s_{Z,1}}[1 - H(e_1^{PZ})]  \nonumber\\
&- {\lambda _{EC}}
-6{{\log }_2}\frac{{17}}{{{\varepsilon _{\sec }}}} -{{\log }_2}\frac{2}{{{\varepsilon _{cor}}}}\rfloor.
\end{eqnarray}
Then the secret key rate \(R\) is ${l \mathord{\left/
 {\vphantom {l N}} \right.
 \kern-\nulldelimiterspace} N}$, where \(N\) is the number of pulses sent by Alice.
\(s_{Z,0}\) \(s_{Z,1}\) and \(e_{1}^{PZ}\) are, respectively, the number of vacuum events, the number of single-photon events and the \textit{phase error rate} associated with single-photon events in \(Z\) basis.
The values of these three parameters need to be estimated with decoy-state method instead of being measured from the experiment directly.
The estimation formulas of \(s_{Z,0}\), \(s_{Z,1}\) and \(e_{1}^{PZ}\) are the same with the ones in Ref.~\cite{lim2014concise}.

It should be noted that the estimation of the number of single-photon events in $X$ basis \(s_{X,1}\) in our protocol is different from the one in Ref. \cite{lim2014concise}.
From Eq.(\ref{Eq:e1PZ}), one can see that in order to accomplish the calculation of \(e_{1}^{PZ}\), \(s_{X,1}\) needs to be estimated first.
In Ref. \cite{lim2014concise}, \(s_{X,1}\) is estimated by using Eqs. (\ref{Eq:sZ0}) and (\ref{Eq:sZ1}) with statistics from the $X$ basis.
Thus, the bounds \(n_{X,\omega }^ -\), \(n_{X,{\upsilon _1}}^ + \), \(n_{X,{\upsilon _1}}^ -\), \(n_{X,\omega }^ +\) and \(n_{X,\mu }^ +\) should be estimated from the measurement values of \(n_{X,\omega }\), \(n_{X,\upsilon_1 }\) and \(n_{X,\mu }\) in $X$ basis, which leads to 5 error terms \cite{lim2014concise}.
Different from the protocol in \cite{lim2014concise}, only two intensities are used in $X$ basis in our protocol.
Here, we assume that the yields of single-photon state in two bases are equal in asymptotic setting\footnote{This is normally satisfied when all the detectors have the same parameters (dark count rate, detection efficiency and after-pulse probability) and are operating according to specification.}.
Then, in finite-key setting, \(s_{X,1}\) can be estimated from \(s_{Z,1}\) by using the random-sampling theory (without replacement) \cite{korzh2015provably} and the result is shown in Eq. (\ref{Eq:sX1}). In this case, only 1 error term arises when \(s_{X,1}\) is estimated.

Such a change of the estimation method of \(s_{X,1}\) causes a minor modification on the secret key rate formula [Eq. (\ref{Eq:l})].
According to the Eq. (B4) in Ref. \cite{lim2014concise}, $21$ error terms emerge during the secrecy analysis for the efficient three-intensity protocol, where 5 error terms come from the estimation of \(s_{X,1}\).
In our protocol, \(s_{X,1}\) is estimated with \(s_{Z,1}\), where 1 error term needs to be taken into consideration.
Therefore, only $17$ error terms need to be composed into the secrecy parameter \(\varepsilon _{\sec }\) when we apply the same secrecy analysis method in \cite{lim2014concise} to our protocol.
Then, we set each error term to a common value \(\frac{\varepsilon _{\sec }}{17}\) and this value is used in both our secret key rate formula [Eq. (\ref{Eq:l})] and finite-size decoy-state analysis [Eqs. (3, 5, 7, 8)].
Next, we will show how to estimate \(s_{Z,0}\), \(s_{Z,1}\) and \(e_{1}^{PZ}\) in our protocol.

\(s_{Z,0}\) is given by
\begin{equation}\label{Eq:sZ0}
 {s_{Z,0}} \ge {\tau _{Z,0}}\frac{{{\upsilon _1}n_{Z,\omega }^ -  - \omega n_{Z,{\upsilon _1}}^ + }}{{{\upsilon _1} - \omega }},
\end{equation}
where \({\tau _{Z,i}} = \sum\limits_{k \in \{ \mu ,{\upsilon _1},\omega \} } {{{{P_k}{P_{Z|k}}{e^{ - k}}{k^i}} \mathord{\left/
 {\vphantom {{{P_k}{P_{W|k}}{e^{ - k}}{k^i}} {i!}}} \right.
 \kern-\nulldelimiterspace} {i!}}} \) is the probability that Alice sends an \(i\)-photon pulse in $Z$ basis, \(P_{k}\) and \(P_{W|k}\) are the probability to choose intensity $k$ and the conditional probability to choose $W$ basis (\(W \in \{ Z,X\}\)) conditional on $k$, and
 \begin{equation}\label{Eq:stanz}
n_{Z,k}^ \pm  = \frac{{{e^k}}}{{{P_k}{P_{Z|k}}}}[{n_{Z,k}} \pm \sqrt {\frac{{{n_Z}}}{2}\ln \frac{{17}}{{{\varepsilon _{\sec }}}}} ],(k \in \{ \mu ,{\upsilon _1},\omega \} ).
 \end{equation}

\(s_{Z,1}\) can be calculated by

\begin{equation}\label{Eq:sZ1}
{s_{Z,1}} \ge \frac{{{\tau _{Z,1}}\mu [n_{Z,{\upsilon _1}}^ -  - n_{Z,\omega }^ +  - \frac{{\upsilon _1^2 - {\omega ^2}}}{{{\mu ^2}}}(n_{Z,\mu }^ +  - \frac{{{s_{Z,0}}}}{{{\tau _{Z,0}}}})]}}{{\mu ({\upsilon _1} - \omega ) - \upsilon _1^2 + {\omega ^2}}}.
\end{equation}

By using a random sampling without replacement \cite{korzh2015provably}, \(s_{X,1}\) can be obtained by
\begin{equation}\label{Eq:sX1}
  {s_{X,1}} \ge N_{1}^X\frac{{{s_{Z,1}}}}{{N_{1}^Z}} - 2N_{1}^Xg(N_{1}^X,N_{1}^Z,\frac{{{s_{Z,1}}}}{{N_{1}^Z}},\frac{\varepsilon_{sec} }{{17}}),
\end{equation}
where
\(N_{1}^W = N{\tau _{W,1}}{P_W}\),
\(C(x,y,z) = \exp (\frac{1}{{8(x + y)}} + \frac{1}{{12y}} - \frac{1}{{12yz + 1}} - \frac{1}{{12y(1 - z) + 1}})
\),

\(g(x,y,z,\varepsilon ) = \sqrt {\frac{{2(x + y)z(1 - z)}}{{xy}}\log \frac{{\sqrt {x + y} C(x,y,z)}}{{\sqrt {2\pi xyz(1 - z)} \varepsilon }}}\)
.

The number of bit errors ${v_{X,1}}$ associated with the single-photon events in $X$ basis is given by
\begin{equation}\label{Eq:vx1}
{v_{X,1}} \le {\tau _{X,1}}\frac{{m_{X,{\upsilon _2}}^ +  - m_{X,\omega }^ - }}{{({\upsilon _2} - \omega )}},
\end{equation}
where \({\tau _{X,1}} = \sum\limits_{k \in \{ {\upsilon _2},\omega \} } {{P_k}{P_{X|k}}{e^{ - k}}{k}}  \),
\({m_X} = {m_{X,{\upsilon _2}}} + {m_{X,\omega }}\),
\begin{equation}\label{Eq:mxk}
m_{X,k}^ \pm  = \frac{{{e^k}}}{{{P_k}{P_{X|k}}}}[{m_{X,k}} \pm \sqrt {\frac{{{m_X}}}{2}\ln \frac{{17}}{{{\varepsilon _{\sec }}}}} ],k \in \{ {\upsilon _2},\omega \}.
\end{equation}

$e_1^{pz}$ is computed by
\begin{equation}\label{Eq:e1PZ}
e_1^{pz} = \frac{{{c_{Z,1}}}}{{{s_{Z,1}}}} \le \frac{{{v_{X,1}}}}{{{s_{X,1}}}} + \gamma (\frac{\varepsilon_{sec} }{{17}},\frac{{{v_{X,1}}}}{{{s_{X,1}}}},{s_{X,1}},{s_{Z,1}}),
\end{equation}
where
\(\gamma (a,b,c,d) = \sqrt {\frac{{(c + d)(1 - b)b}}{{cd\log 2}}{{\log }_2}(\frac{{(c + d)}}{{cd(1 - b)b{a^2}}})}\).

\section{Numerical simulation}
To make a comparison with the efficient three-intensity protocol, we perform some numerical simulations for the fiber-based QKD system with the system parameters in \cite{lim2014concise}, shown in Table \ref{Tab:1}.
These parameters come from recent decoy-state QKD and single-photon detector experiments \cite{frohlich2013quantum,walenta2012sine}.
More specifically, the intensity of weakest decoy state \(\omega=2\times10^{-4}\) and the misalignment error rate \(e_{mis}=5\times 10^{-3}\) are all from the experiment work in \cite{frohlich2013quantum}.
Bob uses an active measurement setup with two single-photon detectors and they have a detection efficiency \(\eta _B=0.1\), a dark count rate \(p_{dc}=6\times 10^{-7}\), and an after-pulse probability \(p_{ap}=0.04\)
\cite{walenta2012sine}.
The dedicated optical fiber is used for quantum channel and the attenuation coefficient of the fibers is $0.2$ dB/km.
For simulation, \(E_Z\) is set to be the average of the observed error rates in $Z$ basis and the error-correction efficiency $f$ is set to be 1.16.
In practice, the cost of error correction \(\lambda _{EC}\) is the size of the information exchanged during the error correction step.
Regarding the secrecy, we also set \({\varepsilon _{\sec }}\) to be proportional to the key length $l$, that is, \({\varepsilon _{\sec }} = \kappa l\) where \(\kappa \) is a security constant and can be seen as the secrecy leakage per bit in final key.
To reduce the optimization complexity, we set \({P_{Z|\omega }} = {P_Z}\).
The parameters \(\{ \mu ,{\upsilon _1},{\upsilon _2},{P_\mu },{P_{{\upsilon _1}}},{P_{{\upsilon _2}}},{P_Z}\} \) are optimized to maximize the secret key rate.

\begin{table}
\caption{
List of parameters for numerical simulations.
\(p_{dc}\) is the dark count rate;
\(p_{ap}\) is the after-pulse probability;
\(\omega\) is the lowest intensity;
\(\kappa \) is the security constant;
${\varepsilon _{cor}}$ is the probability that shared secret keys are nonidentical;
\(e_{mis}\) is the misalignment error rate;
\({\eta _B}\) is the detection efficiency.
}
\label{Tab:1}
\begin{tabular}{ccccccc}
\hline\noalign{\smallskip}
\textrm{\(p_{dc}\)}&
\textrm{\({p_{ap}}\)}&
\textrm{\(\omega\)}&
\textrm{\(\kappa\)}&
\textrm{${\varepsilon _{cor}}$}&
\textrm{\(e_{mis}\)}&
\textrm{\({\eta _B}\)}
\\
\noalign{\smallskip}\hline\noalign{\smallskip}
$6\times 10^{-7}$ & $0.04$ &$2\times 10^{-4}$ &$10^{-15}$&$10^{-15}$ & $5\times 10^{-3}$ &0.1
\\
\noalign{\smallskip}\hline
\end{tabular}
\end{table}

\begin{figure}
\resizebox{0.43\textwidth}{!}{%
\includegraphics{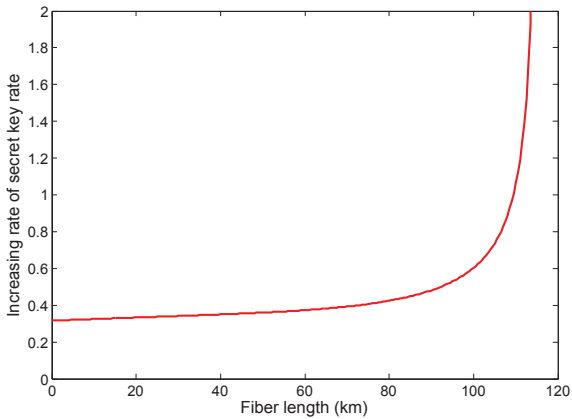}
}
\caption{
 (Color online) The increasing rate of secret key rate vs fiber length (dedicated fiber).
 The red solid line shows the increasing rate of secret key rate between efficient four-intensity protocol and efficient three-intensity protocol.
 This increasing rate is monotonically increasing with the transmission distance.
 }
 \label{Fig:2}
\end{figure}

We set $N$ to be $10^9$ and compare the secret key rates of the efficient three-intensity protocol and our efficient four-intensity protocol.
The results are shown in Fig. \ref{Fig:1}.
Compared with the efficient three-intensity protocol, our protocol can increase secret key rate by at least 30\% at all transmission distances.
The increasing rates of secret key rate at different transmission distances are shown in Fig. \ref{Fig:2}.
We can find that this increasing rate is monotonically increasing with the transmission distance.
That is, the advantage of our efficient four-intensity protocol is more significant at a large transmission distance.
Particularly, at 100 km, the increasing rate is 60\% and the optimal parameters and secret key rates are shown in Table \ref{Tab:2}.
Additionally, we have also performed simulations with other statistical fluctuation analyses, including the standard error analysis \cite{ma2012statistical} and the Chernoff bound \cite{curty2014finite,yin2014long}, and obtain the same conclusion that the secret key rate is increased with our protocol and the improvement is more significant at a large transmission distance.

Compared with the efficient three-intensity protocol, the secret key rate improvement of our efficient four-intensity protocol mainly comes from the following two aspects:

\begin{figure}
\resizebox{0.45\textwidth}{!}{%
\includegraphics{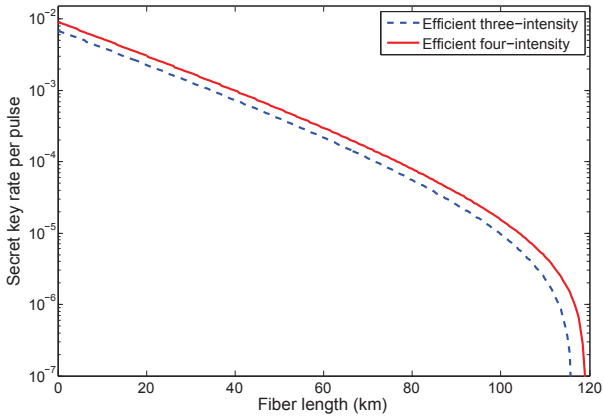}
}
\caption{
 (Color online)
 Secret key rate vs fiber length (dedicated fiber).
 Numerically optimized secret key rates are obtained for a fixed number of pulses sent by Alice $N=10^9$.
 The optimal parameters at the transmission distance of 100 km are shown in Table \ref{Tab:2}.
 The blue dashed line shows the secret key rates of efficient three-intensity protocol \cite{lim2014concise}.
 The secret key rates of efficient four-intensity protocol are presented by the red solid line.
Compared with the efficient three-intensity protocol, our efficient four-intensity protocol can increase the secret key rate by at least 30\%.
Particularly, this improvement is more significant when the transmission distance is large.
The increasing rates of secret key rate at different transmission distances are shown in Fig.~\ref{Fig:2}.
 }
 \label{Fig:1}
\end{figure}

\begin{table}
\caption{
Comparison of parameters at 100 km (standard fiber) between efficient three-intensity protocol \cite{lim2014concise} and our efficient four-intensity protocol.
More general comparison results are shown in Fig. \ref{Fig:1}.
The second and third columns are, respectively, optimal parameters for efficient three-intensity protocol and efficient four-intensity protocol.
Compared with the efficient three-intensity protocol at 100 km, our efficient four-intensity protocol can increase key rate by 60\%.
}
\label{Tab:2}%
\begin{tabular}{ccc}
\hline\noalign{\smallskip}
&
\textrm{Efficient }&
\textrm{Efficient }\\
\textrm{Parameters}&\textrm{three-intensity}&
\textrm{four-intensity}\\
\noalign{\smallskip}\hline\noalign{\smallskip}
{$\mu$}& {$0.551$}& {$0.47$}\\
{$\upsilon_1$}& {$0.188$}& {$0.183$}\\
{$\upsilon_2$}& {$--$}& {$0.32$}\\
{$P_{\mu}$}& {$0.127$}& {$0.16$}\\
{$P_{\upsilon_1}$}& {$0.599$}& {$0.407$}\\
{$P_{\upsilon_2}$}& {$--$}& {$0.22$}\\
{$P_Z$}& {$0.669$}& {$0.82$}\\
{$R$}& {$9.58\times10^{-6}$}& {$1.53\times10^{-5}$}\\
\noalign{\smallskip}\hline
\end{tabular}
\end{table}

(I) In efficient three-intensity protocol, the intensity \(\upsilon _1\) in $Z$ basis and the intensity \(\upsilon _2\) in $X$ basis are replaced by the same intensity \(\upsilon \), and this intensity \(\upsilon\) participates in the calculations of both $s_{Z,1}$ [Eq. (\ref{Eq:sZ1})] and $v_{X,1}$ [Eq. (\ref{Eq:vx1})].
In asymptotic setting, the optimal estimations of $s_{Z,1}$ and $v_{x,1}$ are obtained when the intensity \(\upsilon\) is infinitesimal \cite{ma2005practical}.
Nonetheless, in finite-key setting, with statistical fluctuation [Eqs. (\ref{Eq:stanz},\ref{Eq:mxk})] into consideration, the optimal intensity \(\upsilon\), where the best estimation of $s_{Z,1}$ is achieved, may not help to get a tight estimation of $v_{X,1}$.
Therefore, in our protocol, we make the intensities (\(\upsilon _1\) and \(\upsilon _2\)) independent by adding another intensity.
From Table \ref{Tab:2}, the optimal intensity \(\upsilon _1\) and the optimal intensity \(\upsilon _2\) are, severally, 0.183 and 0.32.
They are quite different.
In a word, the separate optimization of \(\upsilon _1\) and \(\upsilon _2\) helps to achieve a higher key rate.

\begin{figure}
\resizebox{0.48\textwidth}{!}{%
\includegraphics{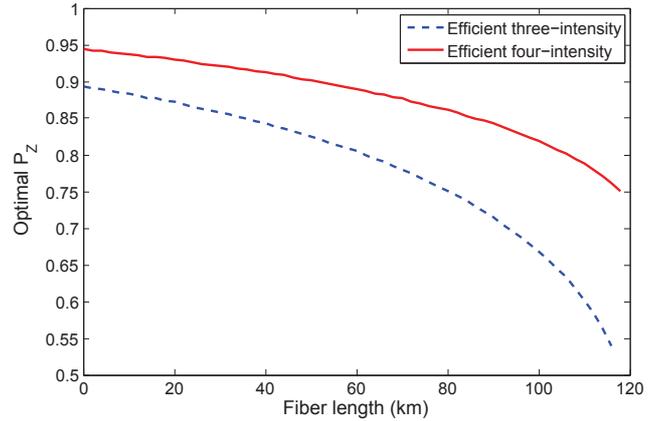}
}
\caption{
(Color online) Optimal $P_Z$ vs fiber length (dedicated fiber).
The blue dashed line shows the optimal $P_Z$s of efficient three-intensity protocol \cite{lim2014concise}.
The optimal \(P_Z\)s of our efficient four-intensity protocol are shown by red solid line.
From the results, we can see that our efficient four-intensity protocol always has a higher \(P_Z\) than the efficient three-intensity protocol.
 }
 \label{Fig:3}
\end{figure}

(II) Compared with standard balanced-basis protocol, efficient protocol can highly improve the secret key rate \cite{lim2014concise,wei2013decoy,lucamarini2013efficient,Lo2005efficient}.
In asymptotic setting, the increasing rate of secret key rate can reach 100\%, when \(P_Z\) approaches 1 \cite{Lo2005efficient}.
However, in finite-key setting, compared to standard three-intensity protocol with balanced basis choice, the efficient three-intensity protocol can increase secret key rate by 45\% \cite{wei2013decoy}.
As the transmission distance increases, the improvement of secret key rate will decrease.
That is because at a larger transmission distance, more pulses in \(X\) basis are needed to give an accurate estimation of $v_{X,1}$ in Eq. (\ref{Eq:vx1}) and Bob has to increase \(P_X\).
When \(P_X\) approaches 0.5, the efficient protocol becomes similar to the standard one, where \(P_Z=P_X=0.5\).
In our protocol, we subtly add an additional intensity and make \(\upsilon _1\) and \(\upsilon _2\) independent.
As Table \ref{Tab:2} shows, we can increase the intensity \(\upsilon _2\) to make the estimation of $v_{X,1}$ more accurate.
That is, with an additional variable to reduce the statistical fluctuation, efficient four-intensity protocol can help to get a higher probability \(P_Z\) to choose \(Z\) basis for key generation.
The optimal \(P_Z\)s for these two protocols at different transmission distances are shown in Fig. \ref{Fig:3}.
From Fig. \ref{Fig:3}, one can see that for these two protocols the optimal \(P_Z\)s are all monotonically decreasing with the transmission distance.
Nevertheless, the optimal \(P_Z\) of our protocol is always larger than the one of efficient three-intensity protocol.
This is the other important factor which leads to a higher secret key rate.

\begin{figure}
\resizebox{0.48\textwidth}{!}{%
\includegraphics{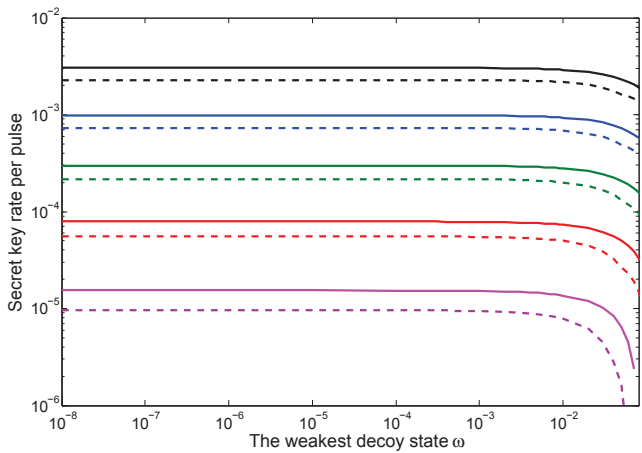}
}
\caption{
(Color online) Secret key rate vs the weakest decoy state \(\omega\).
The optimized secret key rates are obtained for different transmission distances  (20 km, 40 km, 60 km, 80 km, 100km, from top to bottom).
All the solid lines show the results of our efficient four-intensity protocol, e.g., the black solid line shows the optimized secret key rates of our protocol for a fixed transmission distance 20 km.
The results of efficient three-intensity protocol \cite{lim2014concise} are all
presented in dashed lines.
From these results, we find that as long as the intensity \(\omega\) is below \(1\times10^{-3}\), the secret key rates of both the efficient three-intensity protocol and our efficient four-intensity protocol are stable, and our protocol has such an increasing rate of secret key rate as more than 30\% in comparison with the efficient three-intensity protocol at all transmission distances.
}
 \label{Fig:4}
\end{figure}

Note that the intensity \(\omega\) of weakest decoy state in our simulations is set to be \(2\times10^{-4}\) instead 0 (a vacuum state).
That is because, in practice, it is usually difficult to create a perfect vacuum state in decoy-state QKD experiments \cite{rosenberg2007long,dixon2008gigahertz},
although it is optimal to set the weakest decoy state to be vacuum state \cite{ma2005practical}.
In Ref. \cite{lim2014concise}, Lim et al. choose the experiment parameter \(\omega=2\times10^{-4}\) of Ref. \cite{frohlich2013quantum} to perform the numerical simulations.
Following their simulation work, we also set \(\omega=2\times10^{-4}\).
However, what is the effect of the intensity of the weakest decoy state on the secret key rate?
Here, we further optimize the secret key rate over the free \(\omega\).
The results are shown in Fig. \ref{Fig:4}.
We find that as long as the intensity \(\omega\) is below \(1\times10^{-3}\), the secret key rates of both the efficient three-intensity protocol and our efficient four-intensity protocol are stable.
That is, a perfect vacuum state is not essentially required in practical decoy-state QKD experiments.
Meanwhile, we also find that even when the intensity \(\omega\) is free, our protocol can still increase the secret key rate by more than 30\% in comparison with the efficient three-intensity protocol in Ref. \cite{lim2014concise} at all transmission distances.

\section{Discussion and conclusion}

Actually, in terms of current finite-key analysis for decoy-state method, our idea that the intensities and the bases should be independent can be further exploited.
Note that the intensity \(\omega\) also participates in the calculations of both $s_{Z,1}$ in $Z$ basis and $v_{X,1}$ in $X$ basis, and we can add the fifth intensity to make the intensities (\(\omega_1\) in $Z$ basis and \(\omega_2\) in $X$ basis) in two bases independent.
From Eq. (\ref{Eq:e1PZ}), we can see that when the numbers of single-photon events in two bases are close, the sample deviation is small.
Then the sixth intensity, e.g., with an average number of photons of order 1 in $X$ basis, is needed to make the numbers of single-photon events in two bases close.
However, in practical implementations, setting more than four different intensities is hard for experimentalists.
Our efficient four-intensity protocol is feasible and practical for current technology.

In summary, we propose an efficient four-intensity protocol and provide concise finite-key security bounds for this protocol that are valid against general attacks.
Compared with the efficient three-intensity protocol, our efficient four-intensity protocol can increase secret key rate by at least 30\%.
Particularly, at a large transmission distance, the improvement is more significant.

\section*{Acknowledgements}
This work is supported by the National High Technology Research and Development Program of China
Grant No.2011AA010803, the National Natural Science Foundation of China Grants No.61501514 and No.U1204602 and the Open
Project Program of the State Key Laboratory of Mathematical Engineering and Advanced Computing Grant No.2015A13.

\section*{Author contributions}
H.J.,M.G.,B.Y.,W.W.,Z.M. all contributed equally to this paper.

\bibliographystyle{unsrt}
\bibliography{References}

\end{document}